\documentclass[aps,pra,preprint,groupedaddress]{revtex4}
\usepackage{graphicx}

\begin{document}
\title{Adiabatic hyperspherical study of triatomic helium systems}
\author{Hiroya Suno}
\affiliation{Japan Agency for Marine-Earth Science and Technology (JAMSTEC), 
3173-25 Showa-machi, Kanazawa-ku, Yokohama 236-0001, Japan}
\author{B.D. Esry}
\affiliation{Department of Physics, Kansas State University, Manhattan, 
Kansas 66506, USA}
\date{\today}

\begin{abstract}
The $^4$He$_3$ system is studied using the adiabatic hyperspherical 
representation. We adopt the current state-of-the-art helium interaction potential 
including retardation and the nonadditive 
three-body term, to calculate all low-energy properties of the triatomic $^4$He system. 
The bound state energies of the $^4$He trimer are computed  
as well as the $^4$He$+^4$He$_2$ elastic scattering cross sections, the three-body  
recombination and collision induced dissociation rates at finite 
temperatures. We also treat the system that consists of two $^4$He and 
one $^3$He atoms, and compute the spectrum of the isotopic trimer $^4$He$_2\,^3$He, 
the $^3$He$+^4$He$_2$ elastic scattering cross sections, the rates for three-body 
recombination and the collision 
induced dissociation rate at finite 
temperatures. The effects of retardation and the nonadditive three-body 
term are investigated. Retardation is found to be  
significant in some cases, while the three-body term plays only a minor role for these systems.
\end{abstract}

\pacs{}
\maketitle

\section{Introduction}
Three-body atomic systems have attracted considerable interest due to the 
possibility of observing an intriguing property generally referred to as 
"Efimov physics"~\cite{Efimov1,Efimov2,Efimov3}. The most dramatic 
manifestation of Efimov physics is the possibility of an infinity of 
three-body bound states even 
when none exist for the separate two-body subsystems. This occurs 
when the two-body scattering length $a_{12}$ is large compared 
to the characteristic range $r_0$ of the two-body interaction
 potential, which is typically on the  
order of tens of Bohr radii. It is predicted that, in the limit of 
large scattering length (therefore zero two-body binding energy), 
the atom-dimer scattering length also becomes large and the 
three-body system acquires an infinite series of bound states (called 
Efimov states) whose 
energies form a geometric sequence with zero as an accumulation 
point. These phenomena persist even when $r_0/a_{12}$ is nonzero, 
so that we can expect to observe Efimov physics in a real physical 
system. The theory of Efimov physics was formulated in 
1970~\cite{Efimov1,Efimov2,Efimov3}, but was 
experimentally confirmed only in 2006~\cite{Kraemer}, by
exploiting a Feshbach resonance in an ultracold gas 
of $^{133}$Cs atoms to adjust the scattering length. However, the 
evidence of Efimov physics was seen in the 
three-body recombination rates, not by direct observation of bound 
states. It is therefore clear that  
understanding three-body recombination is crucial in establishing 
the extent to which Efimov physics can be realized in the laboratory.

Helium has long been considered to be one of the most promising 
candidates for seeing Efimov physics 
since the $^4$He dimer has a large 
scattering length (larger than 100 a.u.). The theoretical treatment of 
triatomic $^4$He systems is 
simple compared to other atomic species because there exists only one dimer bound state which 
has
zero orbital angular momentum $l=0$. Several sophisticated helium dimer interaction 
potentials have been developed, of which the most widely used one is probably 
the LM2M2 potential by Aziz and Slaman~\cite{Aziz}. This potential has 
been used to calculate various properties of the helium dimer 
and trimer~\cite{PhysRevA.54.394,Blume,0953-4075-31-18-008,
Lazauskas} as well as their scattering 
observables~\cite{PhysRevLett.83.1566,Roudnev,Motovilov,Sandhas}. 
The HFD-B3-FCI1 potential developed by 
Aziz \textit{et al.}~\cite{PhysRevLett.74.1586} has also been used  
to calculate the three-body recombination rates of cold helium 
atoms~\cite{PhysRevA.65.042725,Shepard,braaten-2008}. Jeziorska \textit{et 
al.}~\cite{Jeziorska} and Cencek \textit{et al.}~\cite{Cencek} have recently 
developed not only a helium dimer potential, but also the retardation 
correction to the dimer potential and the nonadditive three-body term. 
Retarded potentials have been used previously~\cite{Sandhas,PhysRevA.64.042514} and 
were found to have clear impacts on the low-energy behavior of helium systems. 
Experimentally, the $^4$He dimer has been observed by 
Luo \textit{et al.}~\cite{Luo1,Luo2} and by Sch\"{o}llkopf and 
Toennies~\cite{Schoellkopf1,Schoellkopf2}. The latter authors were 
able to see not only the helium dimer but also the trimer and 
tetramer. The $^4$He trimer they 
observed was in its ground state, but no experiment has been able 
to see the excited state. There is actually much discussion currently whether 
one or both of these states should be considered Efimov states, 
see Refs.~\cite{Lee,Braaten,Smirne,PhysRevA.67.042706,0953-4075-31-18-008,Roudnev2,
Motovilov,PhysRevA.54.394,PhysRevLett.83.1566,PhysRevLett.82.463,Bedaque,
PhysRevA.60.R9,PhysRevA.67.022505,Bruhl}.

This work is along the line of the investigations in 
Refs.~\cite{PhysRevA.54.394,Blume,0953-4075-31-18-008,PhysRevLett.83.1566,
Shepard,braaten-2008,0953-4075-34-7-101,Roudnev,Motovilov,Sandhas,Kolganova,
PhysRevA.65.042725,Lazauskas}, which dealt with 
the spectrum of the helium trimer, the atom-dimer elastic scattering 
in triatomic helium systems, and the three-body recombination of cold 
helium atoms. None of these previous works, however, presented all of these 
quantities. By doing so, we hope to benefit those applying effective field 
theory to this system. Moreover, we adopt the current state-of-the-art 
triatomic helium interaction 
potential, including the two-body term with retardation 
by Jeziorska \textit{et al.}~\cite{Jeziorska,Szalewicz} and the 
nonadditive three-body term of Cencek \textit{et al.}~\cite{Cencek}. 
Using this complete potential, we compute the bound state energies of the $^4$He trimer 
and the cross sections for elastic scattering between $^4$He and $^4$He$_2$. 
We also calculate the rates for three-body 
recombination $^4$He+$^4$He+$^4$He$\rightarrow ^4$He$_2$+$^4$He and 
collision induced dissociation $^4$He$_2$+$^4$He$\rightarrow ^4$He+$^4$He+$^4$He 
at finite temperatures. In our previous paper~\cite{PhysRevA.65.042725}, we 
published the recombination and dissociation rates using the HFD-B3-FCI1 
potential and including the states with total angular momenta from $J=0$ to 3. 
In the present work, we extend the calculations up to $J=7$ using 
the new potential mentioned above.
In addition, we treat the system consisting of two $^4$He and 
one $^3$He atoms, and calculate both its spectrum and the cross sections 
for elastic scattering between $^4$He$_2$ and $^3$He. We also obtain the rates for 
three-body recombination $^4$He+$^4$He+$^3$He$\rightarrow ^4$He$_2$+$^3$He 
and collision induced dissociation 
$^4$He$_2$+$^3$He$\rightarrow ^4$He+$^4$He+$^3$He at finite temperatures 
for the first time. To summarize, we will present calculations of all low-energy 
properties of triatomic helium for the isotopic combinations that include 
$^4$He$_2$. Moreover, we do so using the best available potentials and examine 
the contribution of the usually-neglected retardation and purely three-body terms.

The key ingredient in our numerical calculations is the adiabatic hyperspherical 
representation~\cite{PhysRevA.54.394,PhysRevA.65.042725,CDLin}. Enforcing the boson 
permutation symmetry is simplified using a modified version 
of the Smith-Whitten coordinate system~\cite{PhysRevA.65.042725,Johnson,Kendrick}. 
The $R$-matrix method~\cite{RevModPhys.68.1015} is used  
to obtain the scattering $S$-matrix. Using this $S$-matrix, we calculate the 
atom-dimer elastic scattering cross sections $\sigma_2$, as well as the rates 
for three-body recombination $K_3$ and collision induced dissociation $D_3$. 
Note that the rate equation for the density of helium atoms in a thermal gas 
can be written as
\begin{equation}
\frac{dn_4}{dt} = -\frac{2}{3!}K_3 n_4^3
-\frac{2}{2!}K_3'n_4^2 n_3 + 2D_3 n_{4d}n_4 +
2D_3' n_{4d}n_3,
\end{equation}
where $n_4$, $n_{4d}$ and $n_3$ are the densities of $^4$He, $^4$He$_2$ and $^3$He, 
respectively. $K_3$ and $K_3'$ are the rates for the recombination processes 
$^4$He+$^4$He+$^4$He$\rightarrow ^4$He$_2$+$^4$He and 
$^4$He+$^4$He+$^3$He$\rightarrow ^4$He$_2$+$^3$He, respectively, while 
$D_3$ and $D_3'$ are the rates for the dissociation processes 
$^4$He$_2$+$^4$He$\rightarrow ^4$He+$^4$He+$^4$He and
$^4$He$_2$+$^3$He$\rightarrow ^4$He+$^4$He+$^3$He, respectively.

This paper is organized as follows. We explain our method and give all 
necessary formulas for calculating the spectra and the collision 
properties of triatomic helium systems in Sec.~\ref{sec:method}. 
The results are presented in 
Sec.~\ref{sec:results}. A summary of this work is given in 
Sec.~\ref{sec:summary}. We use atomic units 
throughout except where explicitly stated otherwise.

\section{Method}
\label{sec:method}
We solve the Schr\"{o}dinger equation for three interacting helium atoms 
using the adiabatic hyperspherical representation 
\cite{PhysRevA.54.394,PhysRevA.65.042725,CDLin}. In the adiabatic hyperspherical 
representation, we calculate eigenfunctions and eigenvalues of 
the fixed-hyperradius Hamiltonian in order to construct a set 
of coupled radial equations. The bound state energies can be 
obtained as the discrete eigenvalues of these coupled equations. 
The scattering $S$-matrix can be extracted from the same coupled 
equations using the $R$-matrix method. The method employed is largely the 
same as detailed in Ref.~\cite{PhysRevA.65.042725}. We thus give here a 
brief outline, emphasizing the modifications necessary when the three particles 
are not identical.

\begin{figure}
\includegraphics[scale=0.6]{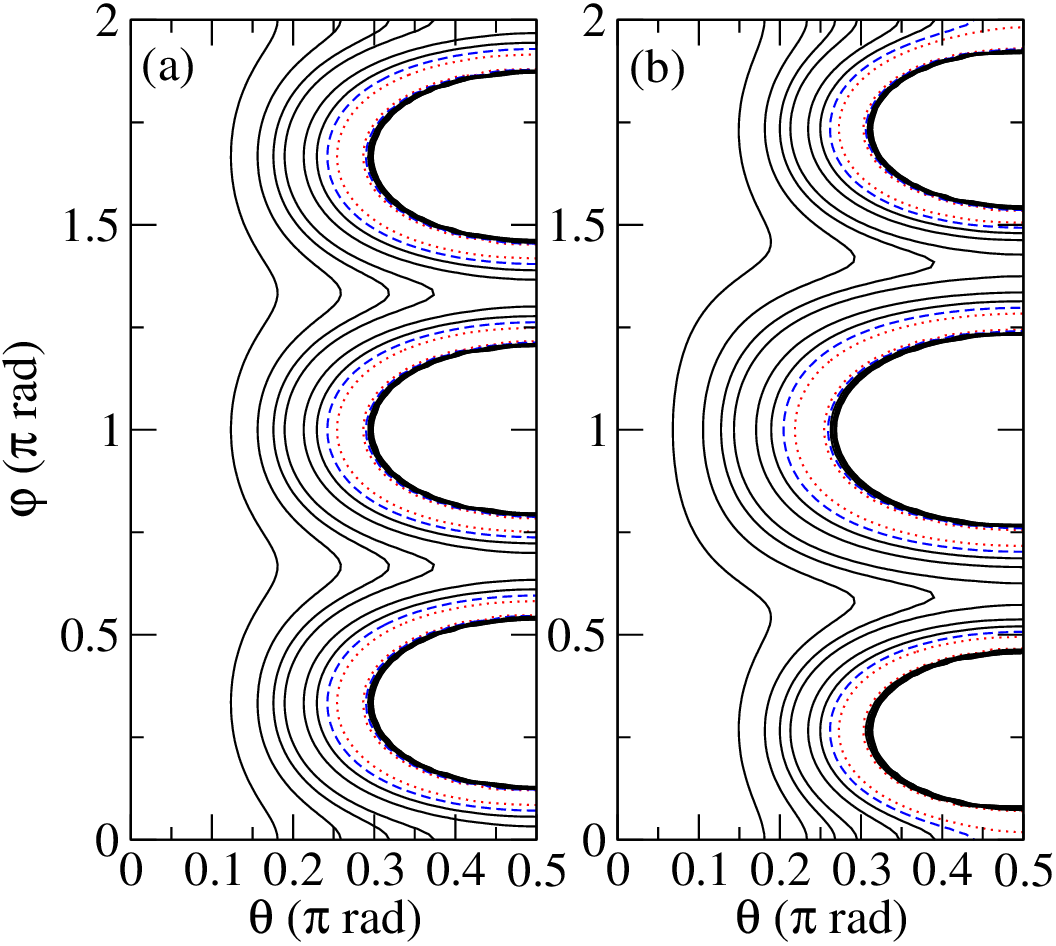}
\caption{(Color online) Contour plot of the potential energy surfaces 
as functions of the hyperangles $\theta$ and $\varphi$ at $R=15$ a.u. for a trimer with 
three identical particles (a) and one with only two identical 
particles (b).
In (a), the translation and reflection symmetries of the potential 
surface can be seen at $\varphi=n\pi/3, (n=1-5)$, while in (b) we 
can identify only the reflection symmetry at $\varphi=\pi$. Exaggerated masses 
are used in (b) to more easily identify the symmetries. The dotted line 
corresponds to the lowest contour line, the dashed line to the second lowest 
contour line.}
\label{Fig1}
\end{figure}

After separation of the center of mass motion, any three-particle system (in 
the absence of an external field) can be described by six coordinates. 
Three of these can be chosen as the Euler angles $\alpha$, $\beta$, 
and $\gamma$ that specify the orientation of the body-fixed frame 
relative to the space-fixed frame. The remaining three internal coordinates 
can be represented by a hyperradius $R$ and two hyperangles $\theta$ and 
$\varphi$. To define these internal coordinates, we use a slightly 
modified version of the Smith-Whitten hyperspherical coordinates 
\cite{PhysRevA.65.042725,Johnson,Kendrick,WS,LPK}. We first introduce 
the three-body reduced mass $\mu$ and the scale parameters $d_{ij}$ as 
follows:
\begin{eqnarray}
\mu^2 & = & \frac{m_1m_2m_3}{m_1+m_2+m_3}, \\
d_{ij}^2 & = & \frac{(m_k/\mu)(m_i+m_j)}{m_i+m_j+m_k},
\end{eqnarray}
where the indices $i$, $j$, $k$ are a cyclic permutation of 
$(1,2,3)$.
The mass-scaled Jacobi coordinates \cite{Delves,Delves2} are defined by 
\begin{eqnarray}
\vec{\rho}_1 & = & (\vec{r}_2-\vec{r}_1)/d_{12},  \\
\vec{\rho}_2 & = & d_{12}\left[\vec{r}_3 - \frac{m_1\vec{r}_1+m_2\vec{r}_2}{m_1+m_2} 
\right].
\end{eqnarray}
Here, $\vec{r}_i$ is the position of the particle $i$ with mass $m_i$ in the lab-fixed 
frame.
We shall write the hyperradius $R$ as follows:
\begin{equation}
R^2 = \rho_1^2+\rho_2^2,\quad R\in[0,\infty).
\end{equation}
The hyperangles $\theta$ and $\varphi$ are defined by 
\begin{equation}
\begin{array}{c}
(\vec{\rho}_1)_{x} = R\cos(\pi/4-\theta/2)\cos(\varphi/2+\varphi_{12}/2), \\
(\vec{\rho}_1)_{y} = R\sin(\pi/4-\theta/2)\sin(\varphi/2+\varphi_{12}/2), \\
(\vec{\rho}_1)_{z} =  0, \\
(\vec{\rho}_2)_{x} =-R\cos(\pi/4-\theta/2)\sin(\varphi/2+\varphi_{12}/2), \\
(\vec{\rho}_2)_{y} = R\sin(\pi/4-\theta/2)\cos(\varphi/2+\varphi_{12}/2), \\
(\vec{\rho}_2)_{z} =  0,
\end{array} \label{Hyperangles}
\end{equation}
where $\varphi_{12} = 2\arctan(m_2/\mu)$. In the case of 
three identical particles, the hyperangle $\varphi$ can be reduced from 
$[0,2\pi]$ to $[0,2\pi/3]$ in which case the interaction 
potential becomes invariant under reflections about $\varphi = \pi/3$. 
Then, in this restricted domain, the solutions of the Schr\"odinger equation 
are automatically either symmetric (bosonic) or antisymmetric (fermionic) 
with respect to exchange of any two particles (this assumes the spin wave 
function is completely symmetric under exchange of any two particles). 
In the case that two of 
the three particles (we choose particles 2 and 3) are identical, the 
interaction potential becomes invariant under reflections about 
$\varphi = \pi$, and the solutions of the Schr\"odinger equation 
are automatically either symmetric (bosonic) or antisymmetric (fermionic) 
in the range $[0,2\pi]$ with respect to exchange of the particles 2 and 3.
Note that the definition of the hyperangle $\varphi$ in 
Eqs.~(\ref{Hyperangles}) is slightly different from that of 
$\varphi^\mathrm{prev.}$ in our previous publication~\cite{PhysRevA.65.042725}.
The relationship between the present and previous definitions is given by  
$\varphi=\varphi^\mathrm{prev.}-4\pi/3$. 
To demonstrate the effect of particle permutations on the 
hyperangle $\varphi$, Fig.~\ref{Fig1} shows contour plots 
of the potential energy surfaces for both three identical particles and 
two identical particles.

In our hyperspherical coordinates, the interparticle distances are given 
by
\begin{equation}
\begin{array}{l}
r_{12} = 
2^{-1/2}{d_{12}R}[1+\sin\theta\cos(\varphi+\varphi_{12})]^{1/2},  \\
r_{23} = 
2^{-1/2}{d_{23}R}[1+\sin\theta\cos\varphi]^{1/2}, \\
r_{31} = 
2^{-1/2}{d_{31}R}[1+\sin\theta\cos(\varphi+\varphi_{31})]^{1/2},
\end{array}
\end{equation}
with
\begin{equation}
\varphi_{12} = 2\arctan(m_2/\mu),\;\varphi_{31} = -2\arctan(m_3/\mu).
\end{equation}

We rewrite the Schr\"odinger equation in terms of a rescaled wavefunction, 
which is related to the usual Schr\"odinger solution $\Psi$ by 
$\psi = R^{5/2}\Psi$. The volume element relevant to integrals over 
$|\psi|^2$ then becomes $2dR\sin 2\theta d\theta d\varphi d\alpha
\sin\beta d\beta d\gamma$. The Schr\"odinger equation for three particles 
interacting through $V(R,\theta,\varphi)$ now 
takes the form
\begin{equation}
\left[- \frac{1}{2\mu}\frac{\partial^2}{\partial R^2}
+ \frac{\Lambda^2}{2\mu R^2} + \frac{15}{8\mu R^2} + V(R,\theta,\varphi) \right]
\psi = E\psi, \label{SchrodingerEquation}
\end{equation}
where $\Lambda^2$ is the squared ``grand angular momentum operator'' 
and its expression is given in Refs. \cite{PhysRevA.65.042725,Kendrick,LPK}.

The interaction potential $V(R,\theta,\varphi)$ 
used in this work is expressed as a sum of three two-body terms 
and a nonadditive three-body term:
\begin{equation}
V(R,\theta,\varphi) = v(r_{12}) + v(r_{23}) + v(r_{31})
+ w(r_{12},r_{23},r_{31}).
\end{equation}
For the helium dimer potential $v(r)$, we use the representation 
of Jeziorska \textit{et al.}, given in Ref.~\cite{Jeziorska}. The 
retardation effect can be incorporated as an additional correction 
term~\cite{Szalewicz} to the dimer potential. Due to this correction, 
the usual van der Waals term $C_6r^{-6}$, valid at short range, transforms 
into $C_7r^{-7}$ at long range. The $^4$He$_2$ bound state energies 
$E_{vl} = E_{00}$ and scattering lengths $a_{12}$ calculated with the 
retarded and unretarded dimer potentials are summarized in 
Table~\ref{Table1} and reproduce those found in Ref.~\cite{Jeziorska}. 
For comparison, the dimer energy with the older LM2M2 potential is 
$-$130 mK~\cite{PhysRevA.54.394}, giving a scattering length of 
100\AA~\cite{Motovilov}. Comparison with other available potentials shows 
general agreement in that all give a weakly bound dimer, but the energy 
can vary by a factor of two~\cite{Motovilov,PhysRevA.64.042514}. There is 
every reason to believe, however, that the current potential is the best 
available (see discussion in Ref.~\cite{Jeziorska}).
No bound state exists for $^4$He$^3$He or $^3$He$_2$.

For the nonadditive three-body term 
$w(r_{12},r_{23},r_{31})$, Cencek \textit{et al.}~\cite{Cencek} give two 
representations. One is based on symmetry-adapted perturbation 
theory (SAPT); the other, on supermolecular coupled-cluster theory with single, 
double and noniterative triple excitations [CCSD(T)]. As mentioned in 
Ref.~\cite{Cencek}, these two representations are very close to each other. 
In test calculations, we find that the three-body problem shows no 
noticeable difference between the SAPT and CCSD(T) representations, 
and the numerical results agree with each other to four significant digits. 
Consequently, the results we present here will use only the SAPT representation 
of the three-body term. In the remainder of this paper, we will use the term 
"complete interaction" to indicate that retardation and the three-body 
term have been included.

The first step that must be carried out is the solution of the fixed-$R$ 
adiabatic eigenvalue equation for a given symmetry $J^\Pi$, with $\Pi$ the 
total parity, to determine 
the adiabatic eigenfunctions (or channel functions) and eigenvalues (or 
potential curves). The adiabatic eigenfunction expansion gives the 
wavefunction $\psi(R,\Omega)$ [we will write $\Omega\equiv(\theta,\varphi,
\alpha,\beta,\gamma)$] in terms of the complete, orthonormal set of 
angular wavefunctions $\Phi_\nu$ and radial wavefunctions $F_{\nu}$
\begin{equation}
\psi(R,\Omega) = \sum_{\nu=0}^{\infty}F_{\nu}(R)\Phi_\nu(R;\Omega).
\label{AdiabaticDecomp}
\end{equation}
The channel functions $\Phi_\nu$ are eigenfunctions of the 
five-dimensional partial differential equation
\begin{equation}
\left[\frac{\Lambda^2}{2\mu R^2}+\frac{15}{8\mu R^2}+V(R,\theta,\varphi)
\right]\Phi_\nu(R;\Omega) = U_\nu(R)\Phi_\nu(R;\Omega), 
\label{AdiabaticEquation}
\end{equation}
whose solutions depend parametrically on $R$. Insertion of 
$\psi$ from Eq.~(\ref{AdiabaticDecomp}) into the 
Schr\"odinger equation from Eq.~(\ref{SchrodingerEquation}) results 
in a set of coupled ordinary differential equations
\begin{equation}
\begin{array}{l}
\left[-{\displaystyle\frac{1}{2\mu}\frac{d^2}{dR^2}}+U_\nu(R)
-\displaystyle\frac{1}{2\mu}Q_{\nu\nu}(R)\right]
F_{\nu}(R) \\
-{\displaystyle\frac{1}{2\mu}}{\displaystyle\sum_{\nu'\neq\nu}}\left[
2P_{\nu\nu'}(R){\displaystyle\frac{d}{dR}}+Q_{\nu\nu'}(R)\right]
F_{\nu'}(R) 
= EF_{\nu}(R).
\end{array} \label{CoupledEquations}
\end{equation}
The coupling elements $P_{\nu\nu'}(R)$ and 
$Q_{\nu\nu'}(R)$ involve partial first and second derivatives of the 
channel functions $\Phi_\nu$ with respect to $R$, 
and are defined as follows
\begin{equation}
P_{\nu\nu'}(R)=\left\langle\!\!\left\langle\Phi_\nu(R;\Omega)
\left|\frac{\partial}{\partial R}\right|\Phi_{\nu'}(R;\Omega)
\right\rangle\!\!\right\rangle,
\end{equation}
and
\begin{equation}
Q_{\nu\nu'}(R)=\left\langle\!\!\left\langle\Phi_\nu(R;\Omega)
\left|\frac{\partial^2}{\partial R^2}\right|\Phi_{\nu'}(R;\Omega)
\right\rangle\!\!\right\rangle. \label{Qmatrix}
\end{equation}
The double-bracket matrix element signifies that integrations 
are carried out only over the angular coordinates $\Omega$.

When solving the adiabatic equation (\ref{AdiabaticEquation}), the degrees 
of freedom corresponding to the Euler angles $\alpha$, $\beta$ and 
$\gamma$ are treated by expanding the channel functions $\Phi_\nu(R;\Omega)$ 
on Wigner $D$ functions~\cite{PhysRevA.65.042725}. The remaining degrees 
of freedom $\theta$ and $\varphi$ are dealt with by expanding 
the channel function onto a direct product of fifth order basis 
splines~\cite{deBoor}. 
We generate the basis splines 
for $\theta$ from 100 mesh points, while we use 80 mesh points 
for $\varphi$. For small hyperradii $R$, a uniform mesh is employed; 
for large $R$, the mesh is designed so that it become dense around 
the two-body coalescence points at 
$(\theta,\varphi)=(\pi/2,\pi-\varphi_{12}), (\pi/2,\pi)$ and 
$(\pi/2,\pi-\varphi_{31})$,
where the potential surface $V(R,\theta,\varphi)$ changes abruptly.
This leads, for example, to a total basis size 
of 34528 in the case of the $3^-$ symmetry. We autoparallelize the 
computer program in order to run it on multiple CPU cores. Typically, 
a calculation of the 32 lowest eigenvalues of the adiabatic 
equation~(\ref{AdiabaticEquation}) takes about 5 minutes of wall clock 
time using 8 1.6-GHz-Itanium2 CPU cores on an SGI Altix 4700 supercomputer. 
We note that our approach, unlike many others employed for this problem, 
does not use a partial wave expansion in the Jacobi-coordinate angular 
momenta. Our results are, in fact, equivalent to including a large number of 
such partial waves.

The identical particle symmetry was built into the adiabatic equations 
via the boundary conditions in $\varphi$. For details, see Ref.~\cite{PhysRevA.65.042725}. 
For the case of three identical particles, 
we have 
\begin{equation}
\Phi_\nu(R;\theta,\varphi =0,\alpha,\beta,\gamma) = 
\pm\Phi_\nu(R;\theta,\varphi =\frac{2\pi}{3},\alpha,\beta,\gamma),\quad\mbox
{for $\Pi=\pm 1$}, \label{BoundaryConditions1}
\end{equation}
and for the case of two identical particles
\begin{equation}
\Phi_\nu(R;\theta,\varphi =0,\alpha,\beta,\gamma) = 
\pm\Phi_\nu(R;\theta,\varphi = 2\pi,\alpha,\beta,\gamma),\quad\mbox
{for $\Pi=\pm 1$}. \label{BoundaryConditions2}
\end{equation}
These conditions ensure that each solution is either symmetric or 
antisymmetric with respect to exchange of any two identical particles 
(about half of the solutions are symmetric), thus eliminating all states 
of mixed symmetry.  
The channel functions with appropriate exchange symmetry are 
then extracted in a postsymmetrization procedure as the solutions 
that satisfy
\begin{equation}
\left\langle\!\!\left\langle\Phi_\nu(R;\Omega)\left|\frac{1+P_{12}}{2}
\right|\Phi_\nu(R;\Omega)\right\rangle\!\!\right\rangle = 1
\end{equation}
for the case of three identical bosons, and 
\begin{equation}
\left\langle\!\!\left\langle\Phi_\nu(R;\Omega)\left|\frac{1+P_{23}}{2}
\right|\Phi_\nu(R;\Omega)\right\rangle\!\!\right\rangle = 1
\end{equation}
for the case of two identical bosons.

In practice we solve the adiabatic equation (\ref{AdiabaticEquation}) 
for a set of about 200 radial grid points $R_i$ up to 
$R\approx 2000\,\mbox{a.u.}$ in order to 
obtain the potential curves $U_\nu(R)$ and the coupling matrix elements 
$P_{\nu\nu'}(R)$ and $Q_{\nu\nu'}(R)$. For 
$R>2000\,\mbox{a.u.}$ they are extrapolated using their known asymptotic 
forms. The details of the numerical calculations are explained in 
Ref.~\cite{PhysRevA.65.042725}.

\section{Results and discussion}
\label{sec:results}
\begin{figure}
\includegraphics[scale=0.5]{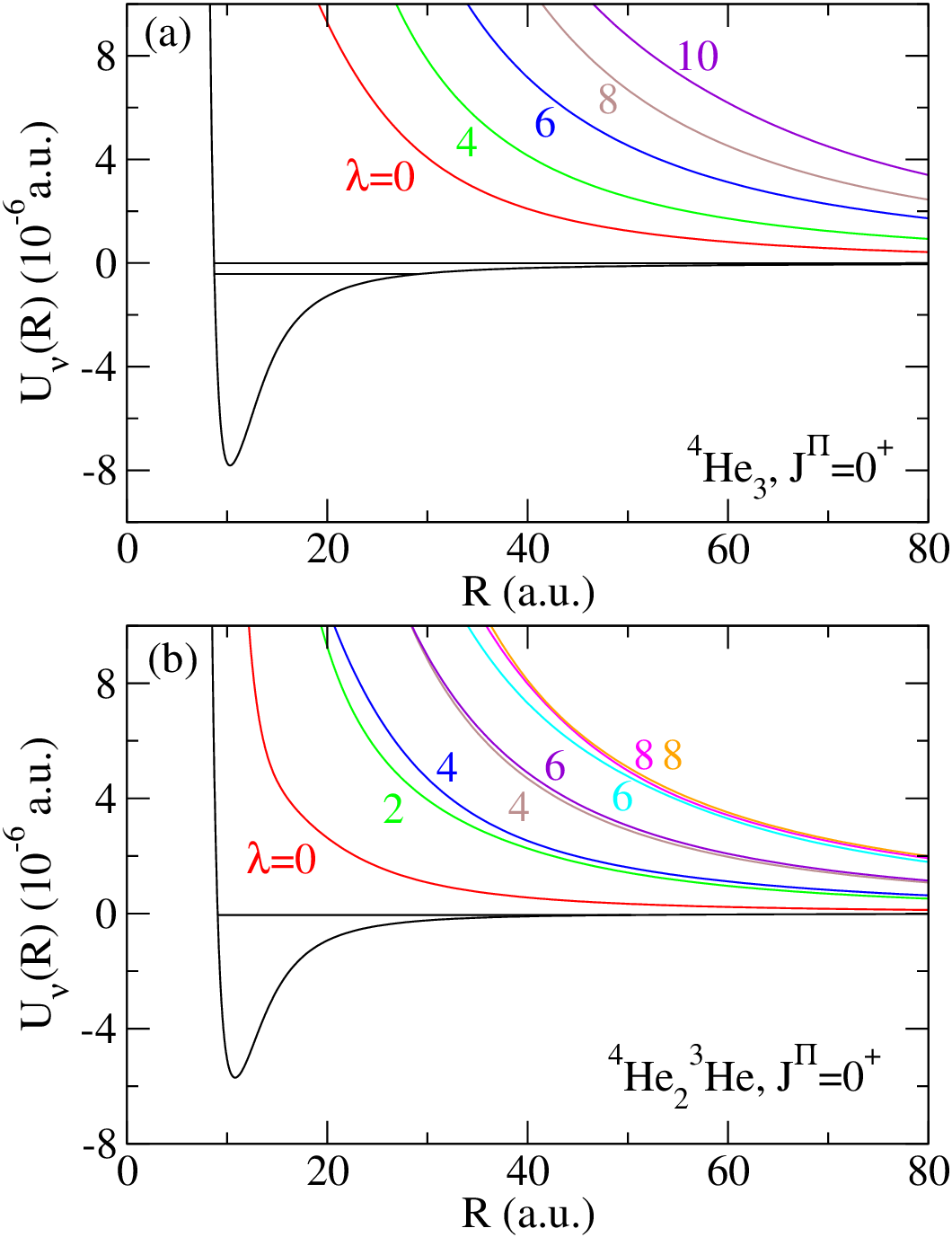}
\caption{(Color online) Adiabatic hyperspherical potential curves $U_\nu(R)$ for (a) 
$^4$He$_3$, $J^\Pi=0^+$ and (b) $^4$He$_2\,^3$He, $J^\Pi=0^+$. The values of 
$\lambda$ indicate the asymptotic 
behavior of the potential curves, as given in Eq.~(\ref{Asymp3BodyChannels}). (a) The 6 lowest 
potential curves $U_\nu(R)$ for 
$^4$He$_3$. In addition, the two bound state energies are shown as horizontal 
lines. (b) The 9 lowest potential curves $U_\nu(R)$ for $^4$He$_2\,^3$He. The bound 
state energy is indicated as a horizontal line.}
\label{Fig2}
\end{figure}

As predicted by several 
authors~\cite{PhysRevA.54.394,Blume,0953-4075-31-18-008,0953-4075-34-7-101,Motovilov,Sandhas,
Kolganova,Lazauskas,Salci,Bressanini,PhysRevA.64.042514}, 
bound states for $^4$He$_3$ and $^4$He$_2\,^3$He 
can exist only for $J^\Pi=0^+$, and none for $J>0$.
Since the $^4$He dimer has only a single $l=0$ bound state, atom-dimer scattering, 
three-body recombination and collision induced dissociation 
are allowed only for the parity-favored cases, that is, $\Pi =(-1)^J$. In the following, 
we therefore limit ourselves to $J^\Pi=0^+,1^-,2^+,$ etc... For each of these 
cases, the lowest adiabatic potential curve $\nu=0$ corresponds asymptotically to two 
atoms bound in a dimer with the third atom far away. Since the dimer has only $l=0$, 
the orbital angular momentum of the relative motion between the dimer and the atom 
should be $l_{1,23}=J$. This potential asymptotically behaves as
\begin{equation}
U_0(R)-\frac{1}{2\mu}Q_{00}(R)\rightarrow E_{00} + \frac{l_{1,23}(l_{1,23}+1)}{2\mu R^2},
\quad\mbox{for $R\rightarrow\infty$}, \label{AtomDimerChannel}
\end{equation}
where $E_{00}$ is the energy of $^4$He$_2$. All the higher 
channels $\nu=1,2,3,...$ for each symmetry correspond to three-body continuum states 
i.e., all three atoms far away from each other as $R\rightarrow\infty$. 
Recall that in the adiabatic hyperspherical representation the three-body 
continuum is rigorously discretized since the adiabatic Hamiltonian 
depends only on the bounded hyperangles. 
These three-body continuum channel functions converge asymptotically to 
the hyperspherical harmonics. Therefore, the corresponding potential 
curves behave as
\begin{equation}
U_\nu(R)\rightarrow\frac{\lambda(\lambda +4)+\frac{15}{4}}{2\mu R^2},
\quad\mbox{for $R\rightarrow\infty$}. \label{Asymp3BodyChannels}
\end{equation}
In principle, $\lambda$ can take on any non-negative integer value, 
but their possible values are restricted by the requirements of 
permutation symmetry. Figure~\ref{Fig2} shows the potential curves 
$U_\nu(R)$ as functions of the hyperradius $R$ for $^4$He$_3$ and 
$^4$He$_2\,^3$He, $J^\Pi=0^+$. Note that we have also calculated 
potential curves for $^4$He$_3$ system with $J^\Pi=1^-,...,7^-$, 
and for $^4$He$_2\,^3$He with $J^\Pi=1^-,2^+$ and $3^-$.

Although we will present cross sections and rates as a function of temperature, 
they are not thermally averaged. Rather, the temperature is obtained from the
energy by the conversion noted in Table~\ref{Table1}. If needed, the thermal 
average can be carried out as described in Ref.~\cite{PhysRevLett.90.053202} 
for $K_3$ or as described in Ref.~\cite{Burke} for two-body processes.

\subsection{Bound state energies}
The bound state energies can be obtained as the 
discrete eigenenergies of the coupled equations in Eq.~(\ref{CoupledEquations}).
We solve these equations by expanding $F_\nu(R)$ onto a set 
of fifth order basis splines in $R$. Typically, a solution of the coupled equations 
including 10 adiabatic channels takes about 1 minute of CPU time using one 
1.6-GHz-Itanium2 processor on an SGI Altix 4700 supercomputer.

The "adiabatic approximation" is the solution of Eq.~(\ref{CoupledEquations}) with only 
one channel $\nu$, which eliminates the sum over $\nu'$ on the left-hand side of 
the equation. This leaves a one-dimensional Schr\"odinger 
equation with an effective hyperradial potential $U_\nu(R)-\frac{1}{2\mu}Q_{\nu\nu}(R)$ 
that determines the three-body spectrum in the adiabatic approximation. 
The lowest $\nu=0$ energy level obtained by solving Eq.~(\ref{CoupledEquations}) 
with all nondiagonal coupling elements neglected is a variational upper bound to the true 
ground state energy $E_0$. One can 
also solve Eq.~(\ref{CoupledEquations}) neglecting the diagonal coupling term 
$Q_{\nu\nu}(R)$. This corresponds to the hyperspherical equivalent of the 
Born-Oppenheimer approximation. The lowest resulting energy for $\nu=0$ gives a rigorous 
lower bound to $E_0$. Finally, solving Eq.~(\ref{CoupledEquations}) with the sums 
truncated at $\nu_\mathrm{max}$ gives variational approximations to the exact energies. 
These energies will thus converge to the exact energies from above in the limit 
$\nu_\mathrm{max}\rightarrow\infty$.

As in previous investigations~\cite{PhysRevA.54.394,Blume,0953-4075-31-18-008,0953-4075-34-7-101,Motovilov,Sandhas,
Kolganova,Lazauskas,Salci,Bressanini,PhysRevA.64.042514}, 
we find two bound states for $^4$He$_3$ and one for $^4$He$_2\,^3$He. 
Table~\ref{Table2} summarizes these different levels of approximation to the bound state 
energies of $^4$He$_3$ and $^4$He$_2\,^3$He as well as the convergence with 
respect to $\nu_\mathrm{max}$. We also include an estimate of the exact energy obtained 
by exprapolating the $\nu_\mathrm{max}$-dependent energies using 
$E_{\nu_\mathrm{max}}=E_\mathrm{exact}+\alpha/\nu_\mathrm{max}^\beta$. 
 The relatively slow convergence in the number of channels 
is due, we think, to the large repulsive core of the helium interaction potential. 
To conclude, we have obtained $E_0=-130.86$ mK and $E_1=-2.5882$ mK for $^4$He$_3$ and 
$E_0=-16.237$ mK for $^4$He$_2\,^3$He with the complete interaction potential.  
For comparisons, Nielsen \textit{et al.}~\cite{0953-4075-31-18-008} obtained 
$E_0=-125.2$ mK and $E_1=-2.269$ mK for $^4$He$_3$ and $E_0=-13.66$ mK for $^4$He$_2\,^3$He, 
using the LM2M2 potential. Using the same LM2M2 potential, 
Blume \textit{et al.}~\cite{Blume} obtained $E_0=-125$ mK and $E_1=-2.28$ mK for $^4$He$_3$. 
Moreover, these energies are also in general agreement with those obtained for other 
two-body potentials.
 
From the table, it can be seen that more than 30 channels are needed to obtain convergence 
to four digits and that $^4$He$_3$ converges more quickly than $^4$He$_2\,^3$He. 
It is also clear that the adiabatic approximation is better than 
the Born-Oppenheimer as has been noted before~\cite{PhysRevA.54.394,Blume}. 
Table~\ref{Table2} also shows the effects of retardation and the three-body terms. 
Comparing the converged results, we see that the complete potential reduces the binding 
compared to the simple pairwise, unretarded potential. In all cases, retardation is more 
important than the three-body term, giving roughly an order of magnitude larger shift in 
the bound state energies. Given the relative importance of retardation, it is clear why 
the energies are shifted upwards since the retardation corrections shortens the tail of 
the potentials from $r^{-6}$ to $r^{-7}$. Physically, it is also easy to understand why 
the three-body term is relatively unimportant once one notes that its leading-order 
contribution is the Axilrod-Teller, dipole-dipole-dipole dispersion term. Because 
helium is closed shell and tightly bound, its polarizability is small, yielding small 
dipersion terms in general. We conclude, then, that the relative importance of 
retardation and the three-body term found here should not be assumed to be a general 
result.

\subsection{Atom-dimer elastic scattering cross sections}
The scattering observables were obtained by solving the 
coupled equations (\ref{CoupledEquations}) using a combination of 
the finite element method (FEM)~\cite{Burke} and the 
$R$-matrix method \cite{RevModPhys.68.1015}. Typically, about 12 adiabatic channels 
(thus $\nu_\mathrm{max}=11$) are used, and $10^4$ elements, in each of which fifth order 
polynomials are used to expand the 
radial wavefunction, extend from $R=5$ to $5\times 10^5\,\mbox{a.u.}$ The scattering 
$S$-matrix is 
then extracted using the $R$-matrix method. 
Each energy took less than 1 minute of wall clock time using one 1.6 GHz 
Itanium2 processor on an SGI Altix 4700 supercomputer. We have checked the 
stability of the $S$-matrix with respect to the final matching distance, 
number of FEM sectors, and the number of coupled channels, and have found our 
results accurate to two significant digits.

The cross section for atom-dimer elastic scattering 
is expressed in terms of the $S$-matrix as 
\begin{equation}
\sigma_2 = \sum_{J,\Pi}\sigma_2^{J\Pi} = 
\sum_{J,\Pi}\frac{(2J+1)\pi}{k_{1,23}^2}|S^{J\Pi}_{0\leftarrow 0}-1|^2.
\end{equation}
Here, $\sigma_2^{J\Pi}$ is the partial atom-dimer elastic scattering cross section 
corresponding to the $J^\Pi$ symmetry, $k_{1,23}=[2\mu_{1,23}(E-E_{00})]^{1/2}$  
is the atom-dimer wavenumber and $\mu_{1,23}$ is the atom-dimer reduced mass 
$\mu_{1,23}=m_1(m_2+m_3)/(m_1+m_2+m_3)$. The event rate constant $K_2$ is related 
to the cross section 
by the formula $K_2=k_{1,23}\sigma_2/\mu_{1,23}$.

The atom-dimer scattering length is another useful quantity that has been 
studied~\cite{Motovilov,Sandhas,Kolganova,Roudnev,Shepard,Lazauskas}.
It is defined as
\begin{equation}
a_{1,23} = - \lim_{k_{12,3}\rightarrow 0}\frac{\tan\delta_0}{k_{1,23}},
\end{equation}
where $\delta_0$ is the phase-shift for atom-dimer elastic scattering and is 
related to the diagonal $S$-matrix element by the formula
\begin{equation}
S^{0+}_{0\leftarrow 0} = \exp(2i\delta_0).
\end{equation}

\begin{figure}
\includegraphics[scale=0.5]{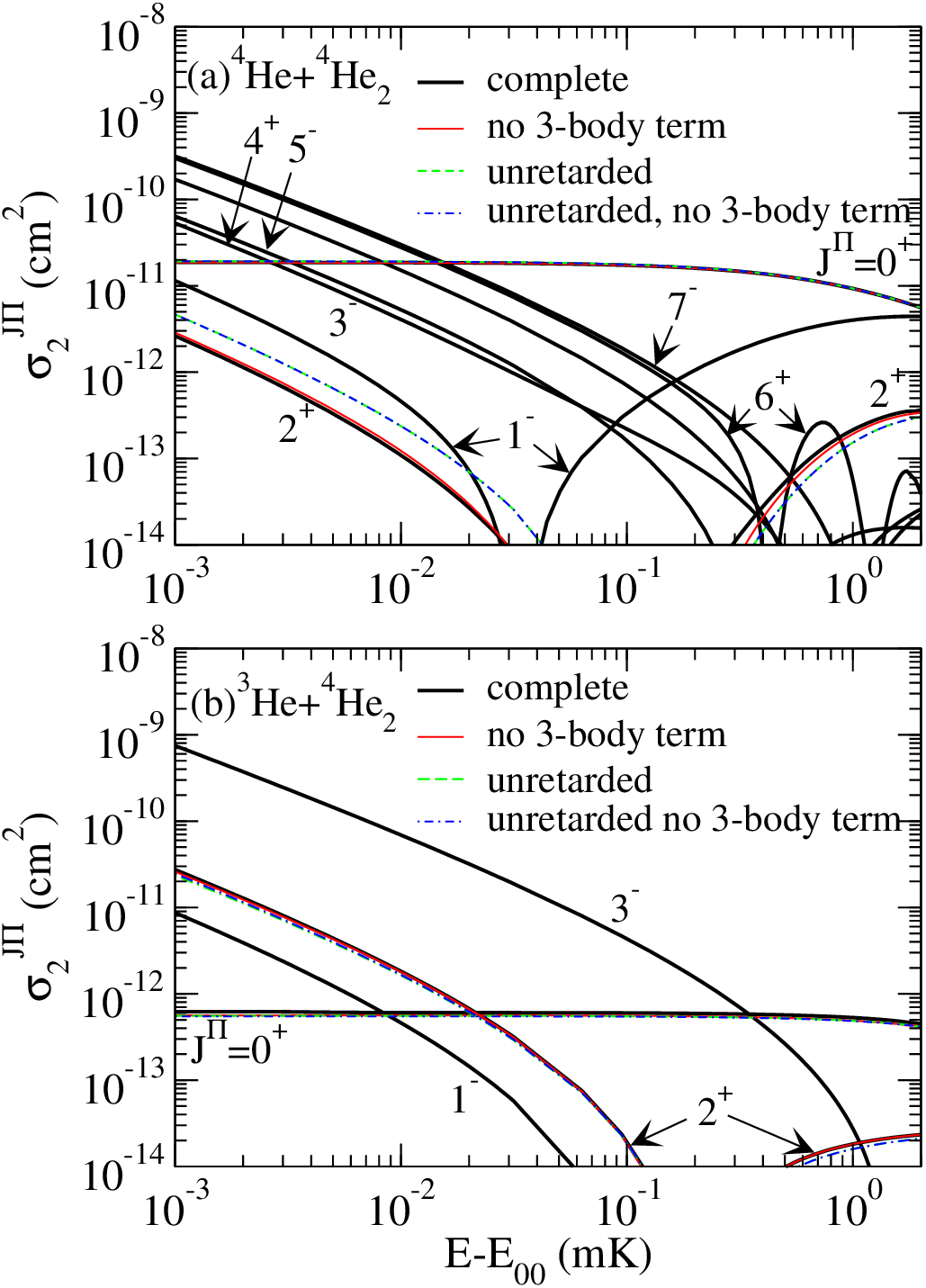}
\caption{(Color online) Partial cross sections $\sigma_2^{J\Pi}$ for (a) $^4$He+$^4$He$_2$ 
elastic scattering in the $J^\Pi=0^+,1^-,$...$,7^-$ symmetries and for (b) 
$^3$He+$^4$He$_2$ elastic scattering in the $J^\Pi=0^+,1^-,2^+$ and $3^-$ 
symmetries, as functions of the collision energy $(E-E_{00})$. The results 
obtained with the 
complete helium interaction potential are shown for collision energies  
$10^{-3}\leq(E-E_{00})\leq 2$ mK, while those calculated 
with various combinations of the retarded two-body potential and three-body term
are also presented for $J^\Pi=0^+$ and $2^+$. Note that the three-body breakup 
threshold is located at $E-E_{00}=1.564$ mK (with the retarded two-body potential) 
or 1.728 mK (with the unretarded potential).}
\label{Fig3}
\end{figure}

Figure~\ref{Fig3}(a) presents the partial cross sections $\sigma_2^{J\Pi}$ for elastic 
$^4$He+$^4$He$_2$ scattering as functions of the collision 
energy $(E-E_{00})$ for the $J^\Pi=0^+,1^-,$...$,7^-$ symmetries. The results obtained 
with the complete helium interaction potential are shown. For comparisons, the figure 
also shows calculations with various combinations of retarded two-body potential and 
three-body term for the $J^\Pi=0^+$ and $2^+$ symmetries. In the ultracold limit, 
here $E-E_{00}<30\mu\mathrm{K}$, $\sigma_2^{0+}$ is constant and tends towards the value 
$\sigma_2^{0+}\rightarrow 4\pi a_{1,23}^2$, with $a_{1,23}=230$~a.u.=120\AA. 
For comparison, Motovilov \textit{et al.}~\cite{Motovilov} obtained 
$a_{1,23}=$115.5\AA, using the LM2M2 
potential. 
Curiously, the $J>0$ partial cross sections decrease as 
functions of the energy for low energy, behaving as $\sigma_2^{J\Pi}\propto (E-E_{00})^{-1}$. 
From the Wigner threshold law, however, one would expect them to behave as 
$\sigma_2^{J\Pi}\propto (E-E_{00})^{J}$, since $J$  
is the orbital angular momentum of the atom relative to the dimer. 
We have found, though, that collision energies of some $\mu$K are still too 
large to recover this threshold behavior and that the cross sections reach the 
threshold regime only at collision energies 
less than about 1 nK. 
The effects of retardation and the three-body term are almost negligible for $J^\Pi=0^+$, 
but for $J^\Pi=2^+$, retardation increases the partial cross section by about 75\% for   
$E-E_{00}< 10^{-2}$ mK, while the three-body term has a smaller effect. The trend is 
reversed at collision energies higher than about 0.3 mK.

In Fig.~\ref{Fig3}(b) we present $\sigma_2^{J\Pi}$ for $^3$He+$^4$He$_2$ elastic 
scattering for the $J^\Pi=0^+,1^-,2^+$ and $3^-$ symmetries. As in Fig.~\ref{Fig3}(a), 
we show the results 
obtained with the complete helium interaction potential and those calculated with all 
combinations of the retarded two-body potential and three-body term. For this system, 
we found $a_{1,23}=40$ a.u.=21\AA. For comparison, Sandhas 
\textit{et al.}~\cite{Sandhas} obtained $a_{1,23}=21.0$\AA, using the LM2M2 potential. 
As found in the $^4$He$+^4$He$_2$ 
scattering, the collision energies 
are found to be still too large to show the Wigner threshold behavior 
$\sigma_2^{J\Pi}\propto (E-E_{00})^{J}$. The effects of retardation and the three-body 
term are almost negligible. 

\begin{figure}
\includegraphics[scale=0.5]{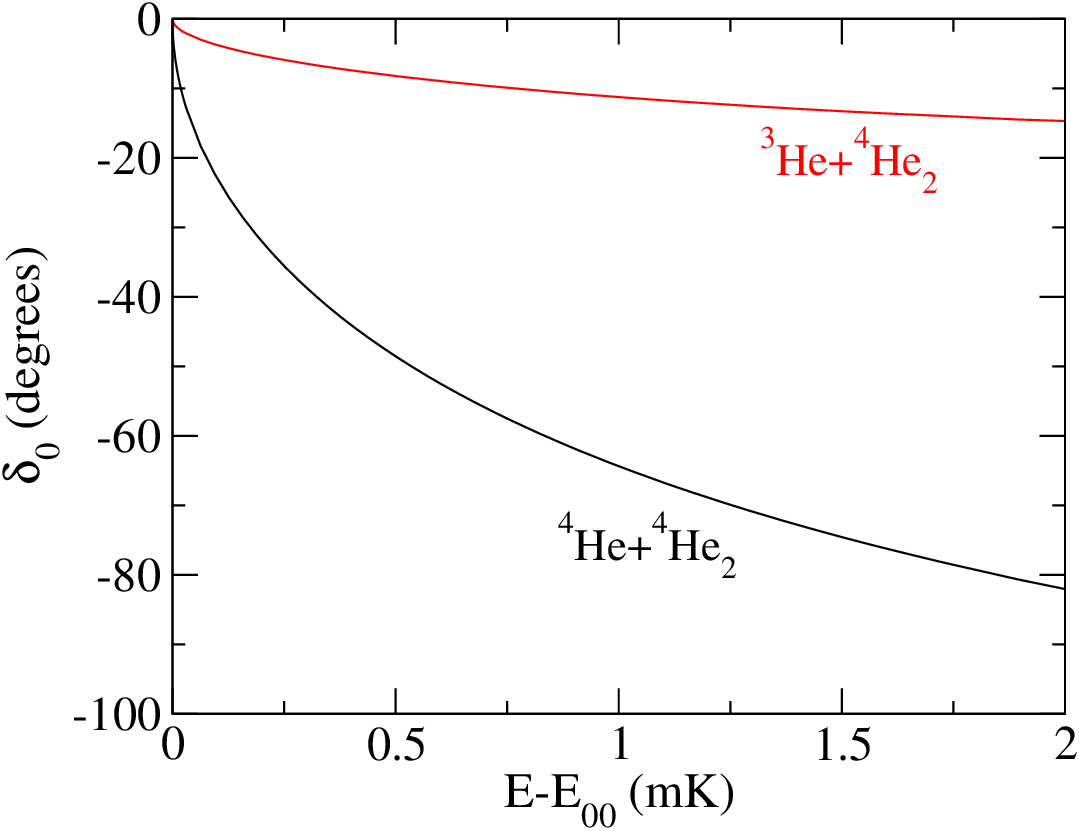}
\caption{(Color online) Phase shifts $\delta_0$ for $s$-wave atom-dimer scattering as functions of the 
collision energy. The lower 
curve shows the results for $^4$He$+^4$He$_2$ and the upper one represents the 
results for $^3$He$+^4$He$_2$. Note that the energies are given relative to the two-body 
dissociation threshold.}
\label{Fig4}
\end{figure}

Figure~\ref{Fig4} shows the phase shifts $\delta_0$ 
for $s$-wave atom-dimer elastic scattering as functions of the collision energy. The lower 
curve shows the results for $^4$He$+^4$He$_2$; and the upper one, for $^3$He$+^4$He$_2$. 
These results were obtained from the complete potential, but we did not observe any 
visible difference from the results obtained neglecting the corrections on this scale. 

\subsection{Three-body recombination rates}
With the $S$-matrix in hand, we can also calculate the three-body recombination rates. 
The event rate constant for three-body recombination for 
$^4$He+$^4$He+$^4$He$\rightarrow ^4$He$_2$+$^4$He is 
\begin{equation}
K_3 = \frac{k}{\mu}\sigma^K_3 = \sum_{J,\Pi}K_3^{J\Pi} = 3! \sum_{J,\Pi}
\sum_{\nu=1}^{\nu_\mathrm{max}}
\frac{32(2J+1)\pi^2}{\mu k^4}
|S^{J\Pi}_{0\leftarrow \nu}|^2. \label{RecombXsec3}
\end{equation}
Here, $K_3^{J\Pi}$ is the partial recombination rate corresponding to the $J^\Pi$ 
symmetry, and $k=(2\mu E)^{1/2}$ is the hyperradial wavenumber in the incident 
channel. $S^{J\Pi}_{0\leftarrow\nu}$ represents scattering from 
the initial three-body continuum channels ($\nu=1,2,...,\nu_\mathrm{max}$) to the 
final atom-dimer channel ($\nu=0$) for the $J^\Pi$ symmetry. The factor (3!) derives from 
the number of indistinguishable bosonic particles. In the same spirit, the rate for 
three-body recombination $^4$He+$^4$He+$^3$He$\rightarrow ^4$He$_2$+$^3$He is given by
\begin{equation}
K_3 = \frac{k}{\mu}\sigma^K_3 = 2!\sum_{J,\Pi}\sum_{\nu=1}^{\nu_\mathrm{max}}
\frac{32(2J+1)\pi^2}{\mu k^4}
|S^{J\Pi}_{0\leftarrow\nu}|^2. \label{RecombXsec2}
\end{equation}
We have checked the stability of the results with respect to the final matching 
distance, number of FEM sectors, and the number of coupled channels. Typically, 
we found the results to be accurate to three significant digits at 100 mK.

\begin{figure}
\includegraphics[scale=0.5]{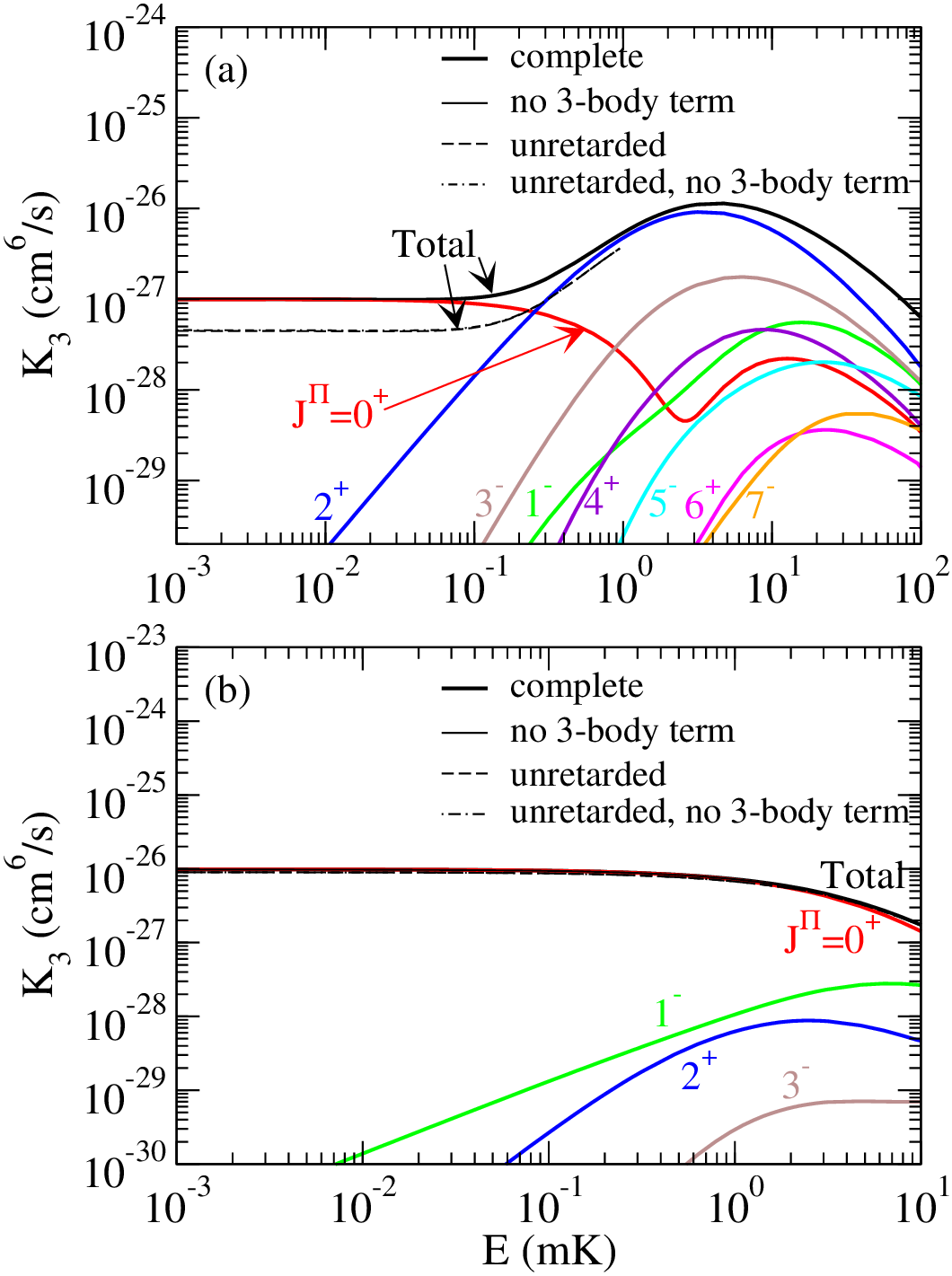}
\caption{(Color online) Total rate $K_3$ and partial rates $K_3^{J\Pi}$ for 
three-body recombination, (a) 
$^4$He+$^4$He+$^4$He$\rightarrow ^4$He$_2$+$^4$He and (b)
$^4$He+$^4$He+$^3$He$\rightarrow ^4$He$_2$+$^3$He, as functions 
of the collision energy $E$. In (a), $K_3$ and $K_3^{J\Pi}$ for the 
$J^\Pi=0^+,1^-,2^+,...,7^-$ symmetries obtained with the complete 
helium interaction potential are shown while $K_3$  
calculated with all combinations of the retarded two-body potential 
and three-body term are also presented in the region $E<1$ mK. In (b), 
$K_3$ and $K_3^{J\Pi}$ for $J^\Pi=0^+,1^-,2^+$ and $3^-$ symmetries 
obtained with the complete helium interaction potential are shown 
along with $K_3$ calculated with all combinations of the retarded 
two-body potential and three-body term.}
\label{Fig5}
\end{figure}

Figure~\ref{Fig5}(a) shows the total rate $K_3$ for three-body recombination 
$^4$He+$^4$He+$^4$He$\rightarrow ^4$He$_2$+$^4$He as well as 
the partial rates $K_3^{J\Pi}$ for the $J^\Pi=0^+,1^-,2^+,...,7^-$ 
symmetries as functions of the collision energy $E$. The results 
obtained with the complete helium interaction potential are shown
while the total rates calculated with the unretarded two-body potential 
and/or the nonadditive three-body term are also presented 
in the region $E<1$ mK. At the lower collision energies, 
$K_3^{J\Pi}\propto E^{\lambda_\mathrm{min}}$, 
where $\lambda_\mathrm{min}$ is the minimum value of $\lambda$ in 
Eq.~(\ref{Asymp3BodyChannels}) allowed by permutation symmetry.
For $J^\Pi=0^+,1^-,2^+$ and $3^-$ in $^4$He$_3$, we have 
$\lambda_\mathrm{min}=0,3,2$ and 3, respectively; for $J>3$, 
$\lambda_\mathrm{min}=J$. These threshold behaviors are predicted by 
a generalized Wigner threshold law~\cite{PhysRevA.65.010705}. 
For collision energies $E<30\mu\mathrm{K}$, the total recombination rate 
$K_3$ is constant, and the $0^+$ partial recombination rate $K_3^{0+}$ dominates. 
In the ultracold limit, we obtain $K_3=9.93\times 10^{-28}$ cm$^6$/s with the 
complete helium interaction potential, while we have $4.45\times 10^{-28}$ cm$^6$/s 
without retardation, $1.01\times 10^{-27}$ cm$^6$/s without the three-body term, and 
$4.54\times 10^{-28}$ cm$^6$/s without either correction. 
Hence, retardation increases the recombination rate by about 120\%, while 
the three-body term has a much smaller effect, lowering the recombination rate by 
about 2\%.
The present results differ from our previous calculations~\cite{PhysRevA.65.042725} 
using the HFD-B3-FCI1 potential, which gave $K_3=7.1\times 10^{-28}$ cm$^6$/s. 
Shepard~\cite{Shepard} also obtained $K_3=7.09\times 10^{-28}$ cm$^6$/s using the 
same potential.

Figure~\ref{Fig5}(b) shows the total rate $K_3$ for three-body recombination 
$^4$He+$^4$He+$^3$He$\rightarrow ^4$He$_2$+$^3$He as well as 
the partial rates $K_3^{J\Pi}$ for the $J^\Pi=0^+,1^-,2^+$ and $3^-$  
symmetries as functions of the collision energy $E$. 
At the lower collision energies, $K_3^{J\Pi}\propto E^{\lambda_\mathrm{min}}$, 
with $\lambda_\mathrm{min}=0,1,2$ and $3$, for $J^\Pi=0^+,1^-,2^+$ and $3^-$, 
respectively~\cite{PhysRevA.65.010705}. 
In the ultracold limit, we obtain 
$K_3=9.83\times 10^{-27}$ cm$^6$/s with the 
complete helium interaction potential, while we have $8.97\times 10^{-27}$ cm$^6$/s 
without retardation, $9.78\times 10^{-27}$ cm$^6$/s without the three-body term, and 
$8.98\times 10^{-27}$ cm$^6$/s without either correction. Hence, retardation 
increases the recombination rate by about 10\%. We observe 
that the three-body term has a smaller effect by far than retardation, lowering 
the recombination rate by only about 0.05\%. Interestingly, recombination is 
much more efficient here with $K_3$ an order of magnitude larger than the homonuclear 
case above.

\subsection{Collision induced dissociation rates}
We can easily calculate the rates for collision induced dissociation since they 
require the same $S$-matrix elements as recombination. The collision induced 
dissociation rate is
\begin{equation}
D_3 = \frac{k_{1,23}}{\mu_{1,23}}\sigma_3^D = \sum_{J,\Pi} D_3^{J\Pi}
=\sum_{J,\Pi}\sum_{\nu=1}^{\nu_\mathrm{max}}\frac{(2J+1)\pi}{\mu_{1,23}k_{1,23}}
|S_{\nu\leftarrow 0}^{J\Pi}|^2,
\end{equation}
where $D_3^{J\Pi}$ is the partial dissociation rate corresponding to the 
$J^\Pi$ symmetry.

\begin{figure}
\includegraphics[scale=0.5]{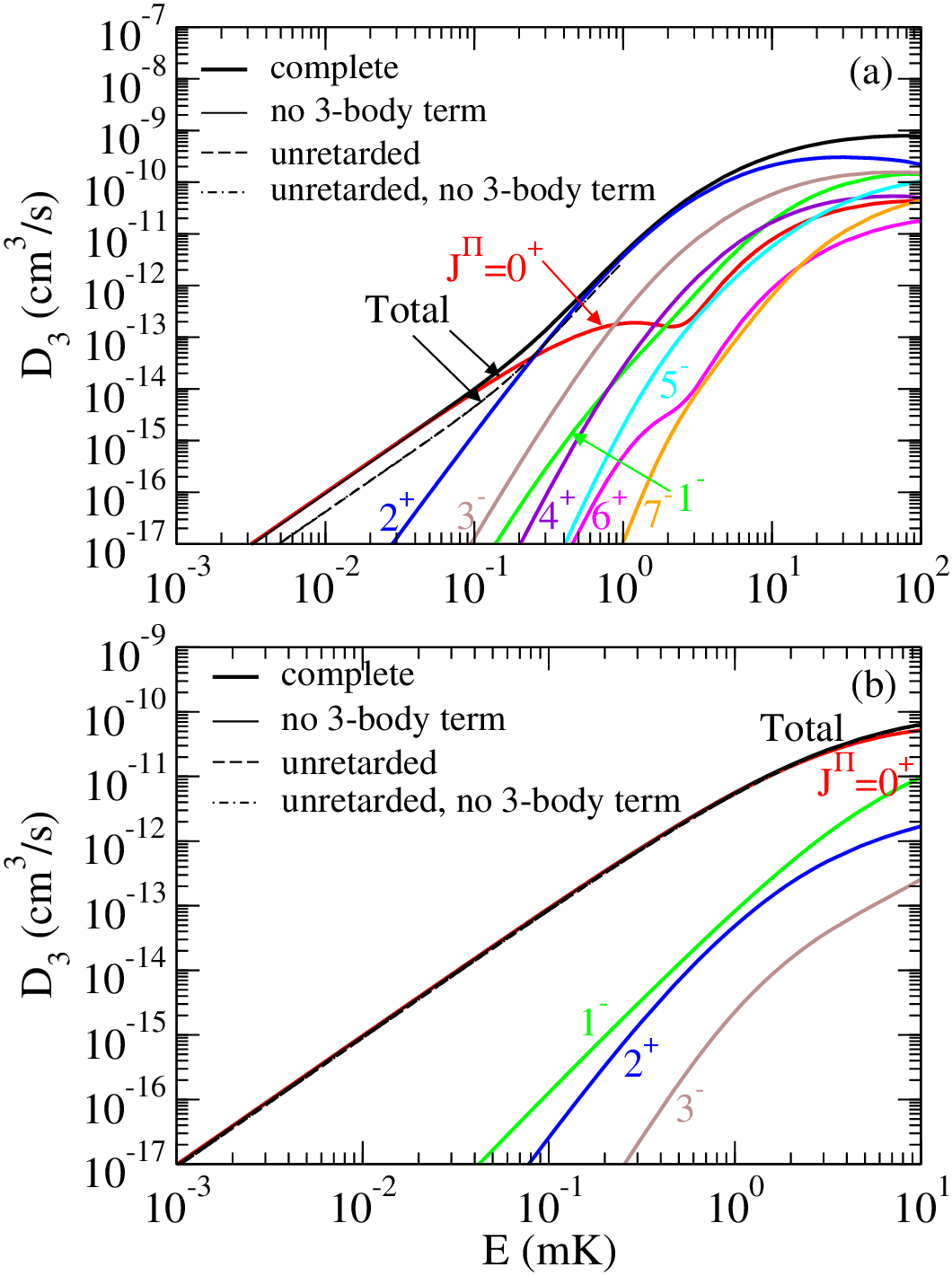}
\caption{(Color online) Total rate $D_3$ and partial rates $D_3^{J\Pi}$ for collision induced 
dissociation, (a) $^4$He$_2$+$^4$He$\rightarrow ^4$He+$^4$He+$^4$He and 
(b) $^4$He$_2$+$^3$He$\rightarrow ^4$He+$^4$He+$^3$He, as functions of 
the collision energy $E$. In (a), $D_3$ and $D_3^{J\Pi}$ for the 
$J^\Pi=0^+,1^-,2^+,...,7^-$ symmetries obtained with the complete 
helium interaction potential are shown while $D_3$  
calculated with all combinations of the retarded two-body potential 
and three-body term are also presented in the region $E<1$ mK. In (b), 
$D_3$ and $D_3^{J\Pi}$ for the $J^\Pi=0^+,1^-,2^+$ and $3^-$ symmetries
obtained with the complete helium interaction potential are shown 
along with $D_3$ calculated with all combinations of the retarded 
two-body potential and three-body term.}
\label{Fig6}
\end{figure}

Figure~\ref{Fig6}(a) shows $D_3$ for $^4$He$_2$+$^4$He$\rightarrow ^4$He+$^4$He+$^4$He 
as well as the partial rates $D_3^{J\Pi}$ for $J^\Pi=0^+,1^-,2^+,...,7^-$. At the 
lower collision energies, $D_3^{J\Pi}\propto E^{\lambda_\mathrm{min}+2}$, with the 
same $\lambda_\mathrm{min}$ as $K_3^{J\Pi}$~\cite{PhysRevA.65.010705}. Because it 
uses exactly the same $S$-matrix elements as $K_3$, the effects of retardation 
and the three-body term are similar. In particular, retardation increases $D_3$ 
by about 134\% at low energies, but the three-body term has little effect, lowering 
the dissociation rate only by a few percent. In Fig.~\ref{Fig6}(b), we show $D_3$ for 
$^4$He$_2$+$^3$He$\rightarrow ^4$He+$^4$He+$^3$He. 

\section{Summary}
\label{sec:summary}
In this work, we have carried out a study of triatomic helium systems using the 
adiabatic hyperspherical approach. Adopting the most realistic, state-of-the-art 
helium interaction potential available, we 
update and extend previous investigations on the bound state and scattering 
properties of the $^4$He$_3$ and $^4$He$_2\,^3$He systems. Based on the most 
accurate helium-helium interactions, these three-body potentials also include 
two-body retardation corrections as well as a non-additive three-body contribution. 
We have thus been able to confirm our statement in Ref.~\cite{PhysRevA.65.042725} 
that the three-body term plays only a minor role. In contrast, the effects of retardation 
are significant, increasing the three-body recombination rate by a factor of two to three.

Systems of ground-state helium atoms are relatively simple because there exists 
only one $l=0$ dimer bound states for $^4$He. Further computational improvements 
must be implemented before we can extend this approach to the more complicated cases 
of alkali atoms, where the larger number of sharp nonadiabatic avoided crossings 
pose difficulties. Further, in alkali systems, we expect that the three-body 
term will play a much larger role as has been recognized 
previously~\cite{PhysRevLett.89.153201,Launay}. 
This work shows, however, that retardation cannot be neglected in calculations of ultracold 
scattering properties based on ab initio potential energies, especially for three-body 
recombination. 

\acknowledgments
We thank Krzysztof Szalewicz for providing us with codes producing the helium 
interaction potential used in this work. BDE acknowledges support from the U.S. 
National Science Foundation and the U.S. Air Force office of Scientific 
Research.

\clearpage

\clearpage

\clearpage

\clearpage

\clearpage

\clearpage

\begin{table}
\caption{$^4$He$_2$ bound state energy $E_{00}$ (1 a.u. = $3.1577465\times 10^8$ mK) and 
scattering length $a_{12}$ calculated with the 
retarded and unretarded dimer potentials.} 
\label{Table1}
\begin{ruledtabular}
\begin{tabular}{ccccc}
potential & $E_{00}$(a.u.) & $E_{00}$ (mK) & $a_{12}$ (a.u.) & $a_{12}$ (\AA) \\ \hline
retarded  & $-4.953\times 10^{-9}$ & $-$1.564 & 173.5 & 91.81 \\
unretarded & $-5.472\times 10^{-9}$ & $-$1.728 & 165.4 & 87.53
\end{tabular}
\end{ruledtabular}
\end{table}

\begin{table}
\caption{Ground and excited state energies for $^4$He$_3$ and ground state energy 
for $^4$He$_2\,^3$He at various levels of approximation. The 
$\nu_\mathrm{max}\rightarrow\infty$ bound state 
energies are obtained by extrapolation.
All energies are given in units of mK, and are relative to the three-body 
break-up threshold.}
\label{Table2}
\begin{ruledtabular}
\begin{tabular}{ccccccccc}
   & \multicolumn{4}{c}{$E_0$} & 
     \multicolumn{4}{c}{$E_1$} \\
retardation            & yes & yes & no  & no & yes & yes & no  & no \\
3-body term            & yes & no  & yes & no & yes & no  & yes & no \\ \hline
$^4$He$_3$             &        &  &  &    &     &     &     &    \\
BO                     & $-$300.86 & $-$301.45 & $-$304.89 & $-$305.48 & $-$3.9277 & $-$3.9315 & $-$4.1839 & $-$4.1879 \\
adiabatic              & $-$109.78 & $-$109.97 & $-$112.06 & $-$112.26 & $-$2.4175 & $-$2.4189 & $-$2.6047 & $-$2.6061 \\
$\nu_{\mathrm{max}}=5$ & $-$129.14 & $-$129.39 & $-$131.70 & $-$131.95 & $-$2.5746 & $-$2.5764 & $-$2.7696 & $-$2.7715 \\
$\nu_{\mathrm{max}}=10$& $-$130.14 & $-$130.39 & $-$132.71 & $-$132.96 & $-$2.5825 & $-$2.5843 & $-$2.7779 & $-$2.7797 \\
$\nu_{\mathrm{max}}=15$& $-$130.44 & $-$130.69 & $-$133.01 & $-$133.27 & $-$2.5848 & $-$2.5866 & $-$2.7803 & $-$2.7821 \\
$\nu_{\mathrm{max}}=20$& $-$130.58 & $-$130.83 & $-$133.15 & $-$133.41 & $-$2.5859 & $-$2.5877 & $-$2.7814 & $-$2.7832 \\
$\nu_{\mathrm{max}}=25$& $-$130.64 & $-$130.89 & $-$133.22 & $-$133.48 & $-$2.5865 & $-$2.5883 & $-$2.7820 & $-$2.7838 \\
$\nu_{\mathrm{max}}=30$& $-$130.68 & $-$130.94 & $-$133.26 & $-$133.52 & $-$2.5868 & $-$2.5886 & $-$2.7824 & $-$2.7842 \\
$\nu_{\mathrm{max}}=35$& $-$130.71 & $-$130.97 & $-$133.28 & $-$133.55 & $-$2.5871 & $-$2.5889 & $-$2.7827 & $-$2.7845 \\
$\nu_\mathrm{max}\rightarrow\infty$ & $-$130.86 & $-$131.12 & $-$133.44 & $-$133.70 & $-$2.5882 & $-$2.5900 & $-$2.7838 & $-$2.7856 \\
$^4$He$_2\,^3$He       &        &  &  &    &     &     &     &    \\
BO                     & $-$90.700 & $-$90.945 & $-$93.201 & $-$93.454 &  &  &  &  \\
adiabatic              & $-$11.088 & $-$11.125 & $-$11.985 & $-$12.024 &  &  &  &  \\
$\nu_{\mathrm{max}}=5$ & $-$14.921 & $-$14.973 & $-$15.982 & $-$16.038 &  &  &  &  \\
$\nu_{\mathrm{max}}=10$& $-$15.432 & $-$15.486 & $-$16.514 & $-$16.572 &  &  &  &  \\
$\nu_{\mathrm{max}}=15$& $-$15.644 & $-$15.699 & $-$16.734 & $-$16.792 &  &  &  &  \\
$\nu_{\mathrm{max}}=20$& $-$15.735 & $-$15.790 & $-$16.828 & $-$16.887 &  &  &  &  \\
$\nu_{\mathrm{max}}=25$& $-$15.818 & $-$15.873 & $-$16.914 & $-$16.973 &  &  &  &  \\
$\nu_{\mathrm{max}}=30$& $-$15.863 & $-$15.918 & $-$16.960 & $-$17.019 &  &  &  &  \\
$\nu_{\mathrm{max}}=35$& $-$15.898 & $-$15.953 & $-$16.996 & $-$17.055 &  &  &  &  \\
$\nu_\mathrm{max}\rightarrow\infty$ & $-$16.237 & $-$16.293 & $-$17.346 & $-$17.405 &  &  &  &
\end{tabular}
\end{ruledtabular}
\end{table}


\end{document}